\documentclass[twocolumn,showpacs,preprintnumbers,amsmath,amssymb,superscriptaddress]{revtex4}

\usepackage{graphicx}
\usepackage{dcolumn}
\usepackage{bm}
\usepackage{epsfig}

\begin{document}

\preprint{PRE/turbulence}

\title{Spectral line shapes in low frequency turbulent tokamak plasmas}

\author{Y. Marandet}
\affiliation{PIIM, Universit\'{e} de Provence, Centre de
St-J\'{e}r\^{o}me F13397 Marseille France}
\author{H. Capes}
\affiliation{DRFC-CEA, 13018 Saint Paul lez Durance Cedex, France}
\author{L. Godbert-Mouret}
\author{M. Koubiti}
\author{R. Stamm}
\affiliation{PIIM, Universit\'{e} de Provence, Centre de
St-J\'{e}r\^{o}me F13397 Marseille France}

\date{\today}

\begin{abstract}
In this paper we investigate the influence of low frequency
turbulence on Doppler spectral line shapes in magnetized plasmas.
Low frequency refers here to fluctuations whose typical time scale
is much larger than those characterizing the atomic processes,
such as radiative decay, collisions and charge exchange. This
ordering is in particular relevant for drift wave turbulence,
ubiquitous in edge plasmas of fusion devices. Turbulent
fluctuations are found to affect line shapes through both the
spatial and time averages introduced by the measurement process.
The profile is expressed in terms of the fluid fields describing
the plasma. Assuming the spectrometer acquisition time to be much
larger than the turbulent time scale, an ordering generally
fulfilled in experiments, allows to develop a statistical
formalism. We proceed by investigating the effects of density,
fluid velocity and temperature fluctuations alone on the Doppler
profile of a spectral line emitted by a charge exchange population
of neutrals. Line wings are found to be affected by ion
temperature or fluid velocity fluctuations, and can in some cases
exhibit a power-law behavior. This study gives some insights in
the appearance of non-Boltzmann statistics, such as L\'{e}vy
statistics, when dealing with averaged experimental data.
\end{abstract}

\pacs{32.70.Jz, 52.35.Ra, 05.40.Fb, 05.40.-a}


\maketitle

\section{Introduction}

Spectral line shape studies have played a major role in the
investigation of the nature of atomic radiators and their
environment, in astrophysics as well as in laboratory plasmas.
Indeed, depending on the dominant line broadening mechanisms, it
is for instance possible to retrieve the electron density or the
ion temperature from the analysis of a given line. However, in
many cases actual plasmas are far from thermal equilibrium, being
inhomogeneous or having non-Maxwellian velocity distributions,
features which significantly complicate the analysis of
experimental data. In addition, these departures from thermal
equilibrium can trigger instabilities, whose growth and non-linear
saturation eventually lead to the onset of turbulence
\cite{Krommes,Garbet01,Horton99}. The importance of investigating
the possible effects that turbulence might have on line shapes has
been acknowledged very early, and the motivations of these studies
were, and still are, two-fold : first there is the need to
quantify the errors introduced by neglecting turbulence in routine
diagnosis based on line-shapes. Then, the possible existence of
significant deviations could be used to diagnose turbulence
itself. An optical diagnostic of turbulence based on passive
spectroscopy would indeed be very convenient, this sustainable
technique being already available on numerous experiments.

Historically, a large number of papers (\cite{HGriem74,Oks} and
references therein)  have dealt with the Stark effect resulting
from turbulent electric fields, such as those associated to a high
supra-thermal level of Langmuir waves \cite{Capes77}. Starting
from the seminal paper by Mozer and Baranger \cite{Mozerbar61},
several models have been devised to include turbulent Stark
broadening in the calculation of line-shapes. The results thus
obtained are relevant to plasmas for which Stark effect is
dominant compared to Doppler effect. There are however situations
for which this ordering is reversed, and important examples are
edge plasmas of magnetic fusion devices such as Tokamaks in the
ionizing regime. For these low density plasmas ($N_{e}\leq 5\times
10^{20}$ m$^{-3}$), Zeeman and Doppler effects are the dominant
broadening mechanisms for low lying lines such as the D$\alpha$
(transition between the levels $n=3$ and $n=2$ of the atomic
deuterium). In such cases, line shape studies essentially provide
measurement of the emitters velocity distribution, and have so far
brought valuable results concerning the origin of neutrals in edge
plasmas \cite{Kubo,Stotler,Hey,Koubiti02}. However, these plasmas
are known to be strongly turbulent, i.e. the level of fluctuation
of the fluid fields characterizing the plasma can rise up to
several tenths of percents \cite{Garbet01}. The experimental
characterization of these fluctuations is of first importance to
analyse drift-wave (DW) turbulence, which is held responsible for
the so called anomalous transport degrading the quality of the
confinement \cite{Horton99}.

As an example, we will consider the case of the Balmer $\alpha$ of
hydrogen isotopes (D$\alpha$ for the case of deuterium), since it
is one of the most routinely monitored line in edge plasmas, being
both intense and optically thin. In sections \ref{sec:two} and
\ref{sec:three}, the expression of the measured line profile is
carefully discussed to emphasize the role of the spatial and time
averages involved in the measure. We show in section \ref{sec:two}
that a neutral population created by charge exchange can be
considered as being in a local equilibrium characterized by the
local density, temperature and fluid velocity of the ions. In
sections \ref{sec:four} and \ref{sec:five}, we will show that in
presence of low frequency turbulence, the Doppler profile gives
access to an apparent velocity distribution. By further developing
the model only briefly presented in \cite{nf2004,europhys05}, this
apparent VDF is reexpressed in terms of the Probability
Distribution Function (PDF) of the fluid fields. In section
\ref{sec:six}, the influence of density, fluid velocity and
temperature are successively investigated in details. Finally, it
is shown in section \ref{sec:seven} that for particular choices of
the statistical properties of the turbulent fluctuations, the
apparent VDF becomes a L\'{e}vy distribution. This result
establishes a clear connection with one of our previous work
\cite{CNSNS}, in which we investigated the possible origin of a
power law behavior observed in the line wings of D$\alpha$ spectra
measured in the former ergodic divertor configuration of the Tore
Supra Tokamak.


\section{\label{sec:two} Expression of the measured spectra}

Let us first define precisely the observable quantity for a
spectrally resolved  passive spectroscopy measurement. First of
all, obtaining the spectrum emitted by the plasma (which will be
referred to as the measured spectrum in the following) from the
raw spectrum involves deconvolution of the apparatus function
$\mathcal{I}_{ap}$. In practice, the theoretical spectrum is
convolved with $\mathcal{I}_{ap}$ before being compared to the raw
spectrum. The radiation emitted by the plasma is integrated both
along the Line Of Sight (LOS) and during the acquisition time of
the spectrometer, denoted by $\tau_{m}$. The observable intensity
$\mathcal{I}_{mes}(\Delta\lambda)$, where $\Delta\lambda$ stands
for the wavelength detuning from the center of the line, is thus
given by the following expression

\begin{equation}\label{eq:start}
\mathcal{I}_{mes}(\Delta\lambda)=\int_{0}^{\tau_{m}}\int_{\mathcal{L}}\mathcal{I}_{loc}(\Delta\lambda,z,t)\frac{\delta
S }{4 \pi z^{2}}dz \ dt ,
\end{equation}

\noindent where $\mathcal{I}_{loc}(\Delta\lambda,z,t)$ is the
local line shape emitted at a given distance $z$ from the detector
along the LOS $\mathcal{L}$. Here $\delta S$ stands for the
detector active area. Assuming the latter to be delimited by
$z_{1}<z_{2}$ such that $L=z_{2}-z_{1}\ll z_{1}$, Eq.
(\ref{eq:start}) reduces to

\begin{equation}
\mathcal{I}_{mes}(\Delta\lambda)\simeq\frac{\delta S }{4\pi
z_{1}^{2}
}\int_{0}^{\tau_{m}}\int_{z_{1}}^{z_{2}}\mathcal{I}_{loc}(\Delta\lambda,z,t)dz
\ dt .
\end{equation}

\noindent We now introduce the local absolute brightness

\begin{equation}
  b(z,t)=\int_{-\infty}^{+\infty}\mathcal{I}_{loc}(\Delta\lambda,
z,t) \ d\Delta\lambda ,
\end{equation}

\noindent and define the local line shape normalized to unity by
$\mathcal{I}(\Delta\lambda,
z,t)=\mathcal{I}_{loc}(\Delta\lambda,z,t)/b(z,t)$. In the
remainder of this paper, we will deal with the measured profile
normalized to unity, given by

\begin{equation}
\mathcal{I}_{mes}(\Delta\lambda)= \frac{1}{\tau_{m}}
\int_{0}^{\tau_{m}}\frac{1}{L}\int_{z_{1}}^{z_{2}}B(z,t)\mathcal{I}(\Delta\lambda,z,t)dz
\ dt .
\end{equation}

\noindent where the relative brightness $B$ is defined by

\begin{equation}
B(z,t)=\frac{b(z,t)}{\frac{1}{\tau_{m}}\int_{0}^{\tau_{m}}\frac{1}{L}\int_{z_{1}}^{z_{2}}b(z,t)dtdz}
\end{equation}

So, the measurement process both entails a spatial
and a time average of the local profile. In order to achieve time
resolved measurements, the acquisition time $\tau_{m}$ should be
chosen shorter than the typical turbulent time, which is associated to the
time variations of the functions $B(z,t)$ and
$\mathcal{I}(\Delta\lambda,z,t)$. However, such a choice would
generally result in spectra having very low signal to noise
ratios. To retrieve
information from time resolved measurements, one must then either
forsake spectral resolution or resort to active techniques, such
as Beam Emission Spectroscopy (BES), which allows to diagnose the
time behavior of turbulent density fluctuations at the edge of
tokamaks (e.g. \cite{Zaslavsky00,Jaku02}). Conversely, if spectral
resolution is needed, the acquisition time should be sufficiently
large so as to ensure reasonable signal to noise ratios. In this
paper we will deal with the latter situation, for the limiting
case in which the acquisition time is much larger than the typical
turbulent time scale. In fact, this generally corresponds to the
actual situation for passive spectral line shapes measurements.
Spectra thus obtained will prove to yield further information than
time resolved experiments. The next step in the modelling consists
in relating the local brightness $B$ and the local profile
$\mathcal{I}$ to the parameters characterizing the plasma.

\section{\label{sec:three} Modelling of the local profile}

In this section we will first describe the model that will be used
to describe the plasma, i.e. a set of fluid equations. The
remainder of the section will present the expressions of the local
brightness and profile relevant to edge plasmas typical
conditions.

\subsection{Plasma description}

We are interested in plasmas which can be described by a set of
$N$ macroscopic fields denoted by
$\mathbf{X}=\{X_{1}(\textbf{r},t),\ldots,X_{N}(\textbf{r},t)\}$,
including the density, temperature and fluid velocity for each
species. These fields are solutions of a set of fluid equations.
For example, the field $X_{i}(\textbf{r},t)$ would satisfy to a
conservation equation

\begin{equation}\label{eq:fluidX}
  \frac{\partial X_{i}}{\partial
t}+\mathbf{\nabla}\cdot\mathbf{\Gamma}=S(\mathbf{X}),
\end{equation}

\noindent where $S(\mathbf{X})$ is a source term and
$\mathbf{\Gamma}$ is a flux, for example given by the sum of a
diffusive and a convective flux $\mathbf{\Gamma}=-\kappa
\mathbf{\nabla}X_{i}+\textbf{u}X_{i}$, $\textbf{u}=X_{j}$ being a
velocity field. It is furthermore assumed that this set of fluid
equations describes a turbulent stationary state, for which the
statistical properties of the plasma do not change during the
acquisition time $\tau_{m}$. For each species, the validity of a
fluid description relies on the ordering $\tau\gg
\nu_{coll}^{-1}$, where $\tau$ is the typical time of variation of
the fluid fields and $\nu_{coll}$ the collision frequency (in the
case of DW turbulence in edge plasmas, we have $\tau\simeq 10-100$
$\mu$s). The fulfillment of this ordering ensures that the VDF
$F_{\gamma}$ of the species $\gamma=i,e$ remains close to a local
Maxwellian at each time and location

\begin{equation}\label{eq:locmaxw}
  F_{\gamma}(\textbf{v},\textbf{r},t)\simeq\sqrt{\frac{m_{\gamma}}{2\pi
  T_{\gamma}(\textbf{r},t)}}\exp\left(-\frac{m_{\gamma}(\textbf{v}-\textbf{u}_{\gamma}(\textbf{r},t))^{2}}{2 T_{\gamma}(\textbf{r},t)}
  \right),
\end{equation}

\noindent where $m_{\gamma}$, $T_{\gamma}(\textbf{r},t)$ and
$\textbf{u}_{\gamma}(\textbf{r},t)$ are respectively the mass, the
temperature field expressed in eV, and the fluid velocity field of
the species labeled by $\gamma$. In the following, we shall
consider the case of a pure deuterium plasma for which
$Z_{eff}=1$.

The calculation of the neutrals VDF requires the use of a refined
model. Indeed, there are different sources of neutrals in edge
plasmas of tokamaks, each of them giving birth to a single class
of neutrals. These classes, characterized by different
temperatures, coexist since the density is usually too low in
order to ensure their complete relaxation toward the background
local equilibrium. The lowest temperature class originates from
the dissociation of molecules released from the wall, whereas
those having larger temperatures are mainly attributed to charge
exchange reactions (e.g. \cite{Kubo,Stotler,Hey,Koubiti02}). In
the following, we will only consider the class of neutrals locally
created by charge exchange reactions, that plays an important role
for the line wings behavior. Indeed, it will be shown that
turbulence essentially affects these regions of the spectra. In
order to model the VDF of these emitters, we can once again take
advantage of the separation of scales between atomic processes and
turbulence. In fact, the inverse of the charge exchange rate is of
the order of a few $\mu$s, i.e. shorter than the typical turbulent
time scale. As a result, the emitters VDF remains at each time
close to that of the ions, given by Eq. (\ref{eq:locmaxw}) with
$\gamma=i$. From the microscopic point of view, the emitter's VDF
thus appears as a Maxwellian characterized by a set of slowly
varying macroscopic fields.

\subsection{The local brightness}

\begin{figure}
\epsfig{scale=0.8 ,file=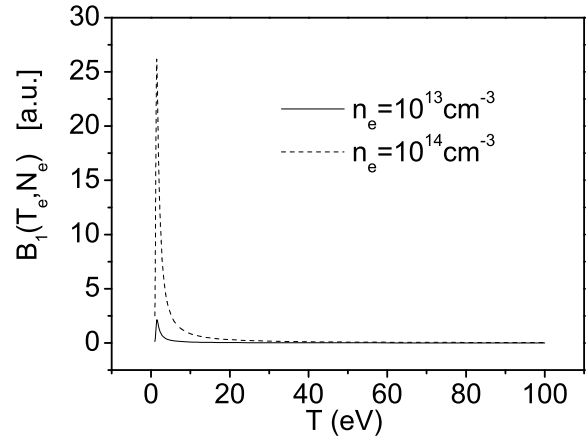} \caption{\label{fig:brightness}
plot of the brightness per emitter as a function of the electronic
temperature $T_{e}$ in the range 0 to 100 eV for a given density
$N_{e}=10^{18}$ m$^{-3}$. The temperature dependence is weak for
$T_{e}>15$ eV.}
\end{figure}

The local line brightness is directly related to the population of
the transition upper atomic level. In general, this population has
to be calculated by taking into account the contributions of the
different processes (for instance collisions, charge exchange,
radiative decay) populating or depopulating the levels.  If the
fluid fields characterizing the plasma vary slowly on the typical
time scales associated to these processes, a stationary approach
is suitable to calculate the brightness. The levels populations
are assumed to be time independent and are calculated using the
values of the fields $\textbf{X}(\textbf{r},t)$ at each time and
location. In practice, the brightness essentially depends on the
electron density $N_{e}(\textbf{r},t)$ and temperature
$T_{e}(\textbf{r},t)$. We have performed a calculation of the
brightness per emitter $B_{1}$ (defined as
$B(\textbf{r},t)=n_{0}(\textbf{r},t)B_{1}(\textbf{X}(\textbf{r},t))$,
where $n_{0}(\textbf{r},t)$ is the density of emitters) for the
$D_{\alpha}$ line in edge plasma conditions, i.e.
$N_{e}=10^{18}-10^{19}$ m$^{-3}$ and $T_{e}=1-100$ eV, using the
code SOPHIA \cite{Rosmej}. The electron density dependence of
$B_{1}$ is found to be linear, in accordance with the fact that
the upper level of the transition is essentially populated by
electronic collisions from the ground state. This leads to a
quadratic behavior of the brightness with $N_{e}$, since
$n_{0}\propto N_{e}$. The influence of the electron temperature on
the brightness is more subtle, as shown on Fig.
\ref{fig:brightness} for two different densities. The existence of
a maximum reflects the competition between the growth of the
electron collisions cross section with temperature, which
dominates the small temperatures behavior, and the ionisation
process. For electron temperatures larger than $15$ eV, the
influence of $T_{e}$ on the brightness is weak, and in the
remainder of this paper, we shall therefore consider the
brightness as being only a function of the electron density.

\subsection{The local line shape}

The $\Delta\lambda$ dependance of the local profile
$\mathcal{I}(\Delta\lambda,z,t)$ is determined by the dominant
line broadening mechanisms. In magnetized plasmas, Zeeman, Stark
and Doppler broadenings should \textit{a priori} be taken
simultaneously into account. In general, the local profile
normalized to unity can be written as the following convolution
product

\begin{equation}\label{eq:completeprofile}
  \mathcal{I}(\Delta\lambda,z,t)=\int d\Delta\lambda' I_{ZS}(\Delta\lambda-\Delta\lambda',z,t)
  I_{D}(\Delta\lambda,z,t) ,
\end{equation}

\noindent where $I_{ZS}$ is the local Zeeman-Stark profile, which
describe the broadening resulting from the effect of the magnetic
and electric fields on the emitters energy levels \cite{HGriem74}.
The Doppler profile $I_{D}$ is related to the wavelength shift
introduced by the movement of the radiator along the LOS, and is
thus directly given by

\begin{equation}\label{eq:Dopl}
  I_{D}(\Delta\lambda,z,t)d\Delta\lambda=f(v_{z},z,t)dv_{z} ,
\end{equation}

\noindent where $f(v_{z},z,t)$ stands for the emitters VDF along
the LOS, obtained from (\ref{eq:locmaxw}) upon integrating over
the two components of the velocity perpendicular to the LOS

\begin{equation}
  f(v_{z},z,t)=\int\int dv_{x}dv_{y}F(\textbf{v},z,t) .
\end{equation}

\noindent It should be noted that Eq. (\ref{eq:Dopl}) would not be
valid if the velocity of the emitter were not constant during the
emission process, due to collisions \cite{Rautian}. If
$\Delta\omega_{D}$ denotes the Doppler line width expressed in
units of pulsation, Eq. (\ref{eq:Dopl}) assumes that
$\tau_{coll}^{-1}\ll \Delta\omega_{D}$. This ordering is largely
satisfied in edge plasmas, and is moreover not inconsistent with
the assumption $\tau_{col}>\tau$ underlying the validity of Eq.
(\ref{eq:locmaxw}). For a given line, the relative importance of
the different broadening mechanisms depends on plasma conditions,
i.e. on the average values taken by the plasma density and
temperature, but also on the detuning $\Delta\lambda$. In the
following, we will again discuss the case of the D${\alpha}$ line,
first for the bulk of the line and then for line wings, these
regions of the spectra for which
$|\Delta\lambda|\gg\Delta\lambda_{1/2}$, $\Delta\lambda_{1/2}$
being the HWHM of the profile. In the center of the line, Stark
effect is negligible for densities lower than $N_{e}=5\times
10^{20}$ m$^{-3}$, an ordering which is usually (but not always)
satisfied in edge plasmas. In addition, since the magnetic field
is larger than $1$ T, fine structure can be neglected
\cite{Bransden}. Therefore, the D$_{\alpha}$ line splits into
three Doppler-broadened Zeeman components (one $\pi$ and two
$\sigma$). The lateral $\sigma$ components are equally separated
from the central $\pi$ component. Under parallel observation with
respect to the magnetic field, only the $\sigma$ components are
observable. Although negligible in the bulk of the line, Stark
effect might become dominant in the line wings for detunings
larger than a value $\Delta\lambda_{S}(N_{e})$ which is an
increasing function of the density. Therefore, in the remainder of
the paper it should be understood that the Doppler line wings are
the regions of the spectra for which both orderings
$|\Delta\lambda|\gg\Delta\lambda_{1/2}$ and
$|\Delta\lambda|<\Delta\lambda_{S}$ are simultaneously valid. The
existence of such a regime depends on the plasma conditions. Its
study is relevant for edge plasmas and consequently Stark effect
will be neglected in the remainder of the paper. However, it
should be emphasized that the statistical formalism which is
developed in section \ref{sec:five} would also be applicable if
Stark effect were not negligible. In the latter case, the local profile should be calculated using Eq. (\ref{eq:completeprofile}) instead of Eq. (\ref{eq:Dopl}).\\
\indent According to Eq. (\ref{eq:Dopl}), the Doppler spectrum of
a single Zeeman component is proportional to the emitters VDF $f$
 along the line of sight. As previously
explained, we consider a class of neutrals created by charge
exchange reactions, whose VDF is approximated by a local
Maxwellian. The corresponding expression of the local Doppler
profile is given by

\begin{widetext}
\begin{equation}\label{eq:Dopplerprofile}
  \mathcal{I}_{D}(\Delta\lambda,T(\textbf{r},t),u_{z}(\textbf{r},t))=\sqrt{\frac{m}{2\pi
  T(\textbf{r},t)}}\exp\left(-\frac{m(\Delta\lambda-\frac{\lambda_{0}}{c}u_{z}(\textbf{r},t))^{2}}{\frac{2\lambda_{0}}{c} T(\textbf{r},t)}
  \right) ,
\end{equation}
\end{widetext}

\noindent where $m$ is the emitters mass, $\lambda_{0}$ the
unperturbed wavelength of the transition under study,
$T(\textbf{r},t)$ the ion temperature, and $u_{z}(\textbf{r},t)$
the component of the ion fluid velocity along the LOS.

\section{\label{sec:four} Apparent velocity distribution}

Gathering the results of the above sections, we obtain the
following expression for the measured profile normalized to unity

\begin{equation}\label{eq:promesurefluide}
  \mathcal{I}_{mes}(\Delta\lambda)=\frac{1}{\tau_{m}}\int_{0}^{\tau_{m}}dt\frac{1}{L}\int_{\mathcal{L}}
  dz  \ B(\textbf{X}(z,t)) I_{D}(\Delta
\lambda,\textbf{X}(z,t)) ,
\end{equation}

\noindent which is now expressed in terms of the fluids fields
describing the plasma. The \textit{apparent
velocity distribution function} $f_{a}(v_{z})$ is straightforwardly deduced from the measured spectrum by

\begin{equation}\label{eq:effVDF}
  \mathcal{I}_{mes}(\Delta\lambda)d\Delta\lambda=f_{a}(v_{z})dv_{z} ,
\end{equation}

\noindent in analogy with Eq.(\ref{eq:Dopl}). This VDF is an average of the local emitters VDF over time and space.
Indeed, combining Eq. (\ref{eq:promesurefluide}) and Eq. (\ref{eq:effVDF}) leads to the following explicit expression

\begin{equation}\label{eq:fbarexpli}
  f_{a}(v_{z})=\frac{1}{\tau_{m}}\int_{0}^{\tau_{m}}dt\frac{1}{L}\int_{\mathcal{L}}
  dz  \ B(\textbf{X}(z,t)) \
f(v_{z},\textbf{X}(z,t)).
\end{equation}

\noindent The apparent VDF $f_{a}$ can be
given a deep physical meaning as will be shown in section
\ref{sec:seven}.\\

\noindent Intuitively, in plasmas where the fluctuation rate is
low, $f_{a}$ should remain close to a Maxwellian $f_{eq}$
characterized by the time and space averaged values of the
temperature and the velocity fields, respectively denoted by
$\bar{T}$ and $ \bar{u}_{z}$, i.e

\begin{equation}\label{eq:avmaxw}
  f_{a}(v_{z})\simeq f_{eq}(v_{z};\bar{T},\bar{u}_{z} ) .
\end{equation}

\noindent Conversely, in a situation where strong fluctuations
occur, there is \textit{a priori} no obvious reason for which the
apparent velocity distribution should remain close to the average
Maxwellian given by Eq. (\ref{eq:avmaxw}). In particular, in edge
plasmas the fluctuation rate can rise up to several tenths of
percents. The validity of Eq. (\ref{eq:avmaxw}) then clearly
becomes questionable, and Eq. (\ref{eq:fbarexpli}) should be used
instead. A calculation of the apparent VDF $f_{a}$ can be carried
out from the latter equation once the solutions of the fluid
equations are known, i.e. the time and space dependences of each
of the fields $X_{i}(\textbf{r},t)$ have been worked out. Due to
the non-linear nature of the fluid equations and the complexity of
the geometry, this calculation would best be achieved numerically.
Although such an approach might be able to encompass the
complexity of the problem, we find it worthwhile to begin with a
simpler one in order to gain insights on the kind of effects that
turbulence might produce on spectral line shapes.


\section{\label{sec:five} Statistical formalism}

\subsection{Expression of the profile}

In the following, we will take advantage of the fact that the
acquisition time of the spectrometer is usually much larger than
the typical time scale of the turbulence $\tau$. Let us first note
that upon using an appropriate normalisation for the $\delta$
function, the following relation holds for any $z$ and $t$

\begin{equation}
  \int_{N}\prod_{i=1}^{N}\delta(\bar{X}_{i}-X_{i}(z,t))d\bar{X}_{1}..d\bar{X}_{N}=1,
\end{equation}

\noindent where $\bar{X}_{i}$ is the sample space variable
corresponding to the field $X_{i}(z,t)$. Introducing this identity
into Eq. (\ref{eq:fbarexpli}), interchanging the order of time and
sample space integrations, and finally making use of the delta
function sifting property yields the following expression for the
apparent velocity distribution function

\begin{widetext}
\begin{equation}
f_{a}(v_{z})=\frac{1}{L}\int_{\mathcal{L}}
  dz \int_{N} d\bar{X}_{1}...d\bar{X}_{N} \left[\frac{1}{\tau_{m}} \int_{0}^{\tau_{m}}
\  \ \prod_{i=1}^{N}\delta(\bar{X}_{i}-X_{i}(z,t))dt\right]
B(\mathbf{\bar{X}}) \ f(v_{z},\mathbf{\bar{X}}).
\end{equation}

\noindent The quantity between brackets is a time average of the
delta functions product, whose typical time variations occur on
the time scale $\tau\ll \tau_{m}$. It is therefore justified to
let $\tau_{m}$ tend to infinity \cite{Frisch95}, and then use the
ergodic assumption, i.e. replace the time average by an ensemble
average denoted by the brackets $\langle\cdot\rangle$

\begin{equation}
\lim_{\tau_{m}\rightarrow +\infty}\frac{1}{\tau_{m}}
\int_{0}^{\tau_{m}} \prod_{i=1}^{N}\delta(\bar{X}_{i}-X_{i}(z,t))
dt =
\left\langle\prod_{i=1}^{N}\delta(\bar{X}_{i}-X_{i}(z,t))\right\rangle
.
\end{equation}
\end{widetext}

\noindent This ensemble average has to be understood as an average
over the time realisations of the stochastic fields $X_{i}(t,z)$
at point $z$, assumed to be a stationary process. Introducing the
local joint Probability Density Function (PDF) of the fluctuating
fields defined by

\begin{equation}\label{eq:definitionPDF}
  \mathcal{P}(\bar{X}_{1},\cdots,\bar{X}_{N},z)=\left\langle\prod_{i=1}^{N}\delta(\bar{X}_{i}-X_{i}(z,t))\right\rangle
,
\end{equation}

\noindent the apparent VDF becomes

\begin{equation}
f_{a}(v_{z})=\frac{1}{L}\int_{\mathcal{L}}
  dz \int_{N} d\bar{\textbf{X}} \mathcal{P}(\bar{\textbf{X}},z)
B(\bar{\textbf{X}}) \ f(v_{z},\bar{\textbf{X}}).
\end{equation}

\noindent Finally, upon integrating on the space coordinate $z$,
the apparent VDF is given by

\begin{equation}\label{eq:profilstats}
f_{a}(v_{z})=\int_{N} d\bar{\textbf{X}} W(\bar{\textbf{X}})
B(\bar{\textbf{X}}) \ f(v_{z},\bar{\textbf{X}}) ,
\end{equation}

\noindent where the spatially integrated PDF $W(\bar{\textbf{X}})$
is obtained from

\begin{equation}\label{eq:integspatW}
  W(\bar{\textbf{X}})=\frac{1}{L}\int_{\mathcal{L}}
  dz \mathcal{P}(\bar{\textbf{X}},z) .
\end{equation}

\noindent In the remainder of the paper we shall furthermore
assume homogeneous turbulence, that is $W(\bar{\textbf{X}})\equiv
\mathcal{P}(\bar{\textbf{X}},z)$ (note that the weaker assumption
of homogeneity along the line of sight is sufficient).



\subsection{Discussion}

In the frame of our statistical reformulation, it is no longer
necessary to know the solutions of the fluid equations in order to
calculate the apparent VDF. Instead, the joint PDF of the
turbulent fields should have been computed. A straightforward
approach would be to rely on a fluid code, so as to compute
histories of the different fields, and then their PDF. As we have
already pointed out, this would require heavy numerical
computation, especially in order to obtain the PDF tails with a
good accuracy. Furthermore, if such calculations were carried out,
any statistical reformulation would obviously be superfluous, and
the apparent VDF could directly be obtained from Eq.
(\ref{eq:fbarexpli}). An approach more suited to our formalism
should proceed directly at the PDF level. The next section will be
devoted to present such a model, initially developed by Pope
\cite{Popebook}. However, it should be emphasized that such a
calculation is bypassed if assumptions for the shape of the PDF
are made. This is the one of the advantages of our formalism,
since it allows to draw conclusions on the properties that
turbulence should have so as to significantly affect line shapes.


\subsection{Determination of the PDF from the fluid equations}

 Let us consider the passive advection of a scalar $X(z,t)$
solution of Eq. (\ref{eq:fluidX}), in which the source term is an
arbitrary function of $X$ and the flux $\mathbf{\Gamma}$ is the
sum of a convective term and a diffusive term. The convective
velocity field $\mathbf{u}$ is assumed to be an incompressible
stochastic field, the statistical properties of which are known.
In order to calculate the apparent velocity distribution function
from Eq. (\ref{eq:profilstats}), the spatially integrated joint
PDF of velocity and temperature, denoted by
$W(\mathbf{u},\bar{X})$, should be calculated. Here, we will limit
ourselves to the modelling of the marginal distribution
$W(\bar{X})$, obtained by integrating $W(\mathbf{u},\bar{X})$ over
the velocity. Indeed, this will be sufficient to highlight the
salient points of the model. Assuming homogeneous turbulence, and
then following Pope \cite{Popebook}, the time dependent PDF
$W(\bar{X},t)$ is shown to obey a Fokker-Planck like equation

\begin{equation}
\frac{\partial W}{\partial
t}=\frac{\partial}{\partial\bar{X}}[S(\bar{X})W]-\frac{\partial^{2}}{\partial\bar{X}^{2}}[D(\bar{X})
W] ,
\end{equation}

\noindent where $S(\bar{X})$ is the source term in the fluid
equation. The expression of the function $D(\bar{X})$ will be
discussed below. The stationary solution of the latter equation is

\begin{equation}\label{eq:solutionstat}
W(\bar{X})=\frac{C}{D(\bar{X})}\exp\left(-\int_{0}^{\bar{X}}\frac{S(w)}{D(w)}dw
\right) .
\end{equation}

\noindent As a result, in the PDF approach a non-linear source
term $S$ does not introduce any closure problem, unlike in the
moment based models \cite{Krommes}. The problem remains
nonetheless unclosed, since the function $D(\bar{X})$ is in
general not expressible in terms of $W(\bar{X})$ or $S(\bar{X})$
alone. Indeed, the shape of this function depends on the
correlations between $X$ and its gradient. More precisely, it can
be recast in the following form

\begin{equation}
  D(\bar{X})=\frac{1}{\langle\mathbf{\nabla}X\rangle^{2}}\int d(\mathbf{\nabla} \bar{X})\mathcal{P}(\mathbf{\nabla} \bar{X} |
  \bar{X})(\mathbf{\nabla} \bar{X})^{2} ,
\end{equation}

\noindent where $\mathcal{P}(\mathbf{\nabla} \bar{X} | \bar{X})$
is the PDF of the gradient of X, conditioned to a given value of X
\cite{Popebook}. In order to obtain this PDF, an equation for the
joint PDF of X and its gradient should be written \cite{Chen89},
which in turn would involve correlations with higher orders
gradients. Eventually, one ends up with an infinite hierarchy of
equations, involving the joint PDFs of $X, \nabla X, \nabla^{2}
X,...$ . In addition, it should be kept in mind that the
statistical properties of the velocity field $\textbf{u}$, while
not appearing explicitly in Eq. (\ref{eq:solutionstat}), do
actually affect the shape of $D(\bar{X})$ through Eq.
(\ref{eq:fluidX}), as should the expression of $S(X)$. The closure
of this hierarchy has proven to be difficult to address. Promising
techniques, such as the \textit{mapping closure} \cite{Chen89}
have been devised to overcome these difficulties, but have not yet
led to decisive results (for an application to the Hasegawa-Mima
equation governing plasma turbulence, see Ref. \cite{Das95}).
Addressing these issues is largely beyond the scope of the present
paper, and for our purposes it will be sufficient to present an
early attempt to this closure problem, due to Sinai and Yakhot
\cite{Sinai89}. These authors were interested in the case of
passive advection of temperature in homogeneous decaying
turbulence, for which there is no source term in the temperature
equation. Their idea is to deal with the rescaled quantity
$X=T/\langle T^{2}\rangle$, which is solution of an equation
analogous to (\ref{eq:fluidX}), $S(\bar{X})$ being a linear
function of $\bar{X}$. The following Taylor development is used to
express the function $D(\bar{X})$

\begin{equation}\label{eq:Tayldev}
D(\bar{X})\simeq 1+k\bar{X}^{2} ,
\end{equation}

\noindent where the parameter $k>0$ is a measure of the
correlations strength. From Eq. (\ref{eq:solutionstat}), the
following result for the temperature PDF is readily obtained

\begin{equation}\label{eq:Tsallisinai}
  W(T)=\frac{C}{\left(1+k\left(\frac{T-T_{0}}{\sigma}\right)^{2}\right)^{1+1/2k}} ,
\end{equation}

\begin{figure}
\epsfig{scale=0.8 ,file=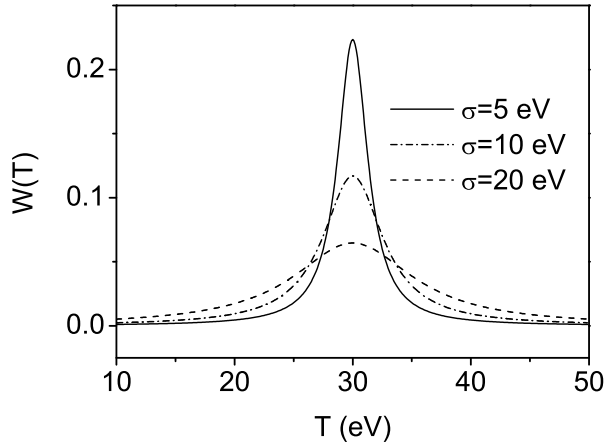} \caption{\label{fig:sinaisigma}
Plot of the Sinai PDF for $T_{0}=30$ eV, $k=10$ and for different
values of $\sigma=5,10,20$ eV. These distributions are used to
compute the corresponding apparent VDF on Fig.
\ref{fig:sinaiprofile}}
\end{figure}

\noindent where $C$ is a normalization constant, and $\sigma$
controls the width of the distribution. Fig. \ref{fig:sinaisigma}
shows a plot of $W(T)$ for $T_{0}=30$ eV, $k=10$ and $\sigma=5, \
10, \ 20$ eV. A few subtleties and limitations concerning the use
of this result deserve to be mentioned. First, it should be noted
that $\sigma$ is actually time dependant. We shall assume here
that the acquisition time $\tau_{m}$ is chosen such that
$\sigma/\dot{\sigma}\ll \tau_{m}$. This requires a separation of
time scales between the turbulent fluctuations and the decay of
the average quantities. Secondly, the correlations are treated
using the development given by Eq. (\ref{eq:Tayldev}), which is
not valid for large values of the temperatures. Our results
concerning line wings should thus be limited to not too large
detuning $\Delta\lambda$. Finally, it should be pointed out that
the distribution given by Eq. (\ref{eq:Tsallisinai}) is a Tsallis
distribution (e.g. \cite{Tsallis95}) with $q=(1+2k)^{-1}$.
Therefore, in this model, temperature fluctuations obey to Tsallis
non-extensive statistical mechanics \cite{Tsallis95} when
correlations exist, and to Boltzmann statistics for vanishing
correlations.


 In the next section, we shall use these
results as an input for apparent VDF calculations.


\section{\label{sec:six} Application to the case of one fluctuating variable}

In an actual turbulent plasma, several fields fluctuate, and these
fluctuations are coupled. According to Eq. (\ref{eq:profilstats}),
the joint PDF of the relevant fields should be computed before
calculating the apparent VDF. However, the role of density,
velocity and temperature fluctuations on the apparent VDF shape
have no reason to be identical. As a first approximation, it is
therefore rational to consider the idealized case in which only
one field fluctuates. This will shed light on which field
fluctuations lead to the most significant effects on line shapes.

\subsection{Density fluctuations}

Let us first consider density fluctuations. Since the local VDF
normalized to unity does not depend on density, the integration
over density fluctuations is trivially performed, and the apparent
VDF is found to be equal to the local emitters VDF

\begin{equation}\label{eq:flucdens}
f_{a}(v_{z})=\int_{0}^{+\infty} dn \ B(n)\ W(n) \
  f(v_{z},T)=f(v_{z},T) .
\end{equation}

\noindent Therefore, at this level of approximation, Doppler line
shapes are not sensitive to density fluctuations. The apparent VDF
should thus remain Gaussian with the temperature $T$, whatever the
shape of $W(n)$. This is in sharp contrast with line brightness
time resolved measurements, which essentially provide information
on density fluctuations. However, it should be noted that for
cases in which Stark effect is not negligible, Eq.
(\ref{eq:flucdens}) no longer holds, since the local line shape
then strongly depends on the density. As we have already pointed
out, the formalism presented here could nevertheless be used upon
replacing the local Doppler profile by the total profile given by
Eq. (\ref{eq:completeprofile}).


\subsection{Fluid velocity fluctuations}

Let us now investigate the case in which only the fluid velocity
fluctuates. In the following, $W(u_{z})$ stands for the PDF of the
fluid velocity component along the line of sight, and
$\sigma_{u}^{2}$ for its variance. Starting from Eq.
(\ref{eq:profilstats}), the apparent VDF reduces to

\begin{equation}\label{eq:fluctuvelocity}
  f_{a}(v_{z})=
\int W(u_{z})f(v_{z}-u_{z},T) \ du_{z} ,
\end{equation}

\noindent which is the convolution product of $W$ and the local
Maxwellian. Eq. (\ref{eq:fluctuvelocity}) is a well known result
in plasma spectroscopy, which is mentioned in classical textbooks
\cite{Griem97}. A shape-independent definition of the apparent
temperature $T_{eff}$ from the profile should proceed from its
second moment

\begin{equation}\label{eq:Teff}
  \xi
  T_{a}=\int_{-\infty}^{+\infty}f_{a}(v_{z})v_{z}^{2}dv .
\end{equation}

\noindent In the fluctuations-free case, the actual temperature of
the emitters $T$ is recovered, whereas if fluctuations do occur
the apparent temperature is given by

\begin{equation}
T_{a}=T\left[1+\frac{\sigma_{u z}^{2}}{v_{th}^{2}}\right] ,
\end{equation}

\noindent where $v_{th}$ is the thermal velocity corresponding to
the temperature $T$. The apparent temperature obtained from the
Doppler line width is thus not rigorously equal to the actual
temperature of the emitters. This result has already been
mentioned by several authors, and was actually used in the first
models retaining the effect of turbulence on Doppler line shapes
\cite{Unsoeld}. In order to obtain a ten percents discrepancy
between $T_{a}$ and $T$ for deuterium emitters, the fluctuation
rate should be of the order of thirty percents (i.e.
$\sigma_{u}\sim 0.3 \ v_{th}$). This effect would be stronger for
heavy emitters, since their thermal velocity is smaller
\cite{Griem97}. This estimation suggests that the $D_{\alpha}$
line width is not strongly modified by fluid velocity
fluctuations. However, considering only the line width is not
sufficient. In fact, the line shape, i.e. the apparent VDF, is
often found to be non-Gaussian, and therefore from Eq.
(\ref{eq:fluctuvelocity}), so should be the PDF $W(u_{z})$. Recent
findings in astrophysical spectra
\cite{Iganov03,Ding99,YingLiu02}, as well as in tokamak plasmas
for radial velocity fluctuations \cite{Jha03} indicate strong
deviations from the Maxwellian, especially for line wings. As an
illustration, let us consider PDFs which have power law tails
characterized by an exponent $\alpha$ such that $0<\alpha<2$. It
is easily shown from Eq. (\ref{eq:fluctuvelocity}) that the
resulting apparent VDF features a similar asymptotical dependence

\begin{equation}
  f_{a}(v_{z})\propto
\frac{1}{|v_{z}|^{\alpha+1}} .
\end{equation}

\begin{figure}
\epsfig{scale=0.8 ,file=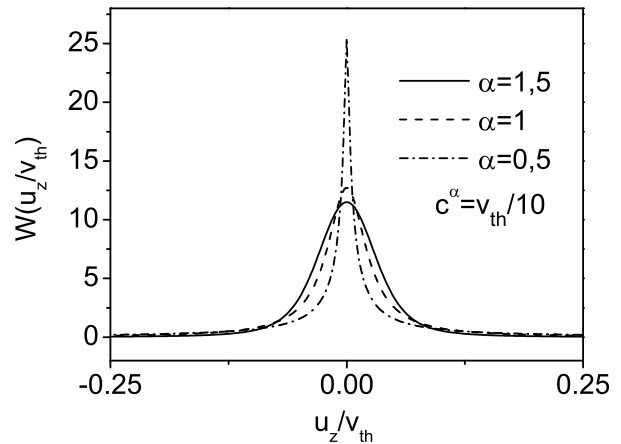}
\caption{\label{fig:levyvelocity} Plot of the L\'{e}vy velocity
PDF for $\alpha=0.5,  \ 1,\ 1.5 $. The fluid velocity is plotted
in units of the thermal velocity $v_{th}$. The parameter $c$
characterizing the width of the distribution (see section
\ref{sec:five}) is defined by $c^{\alpha}=v_{th}/10$.}
\end{figure}

\begin{figure}
\epsfig{scale=0.8 ,file=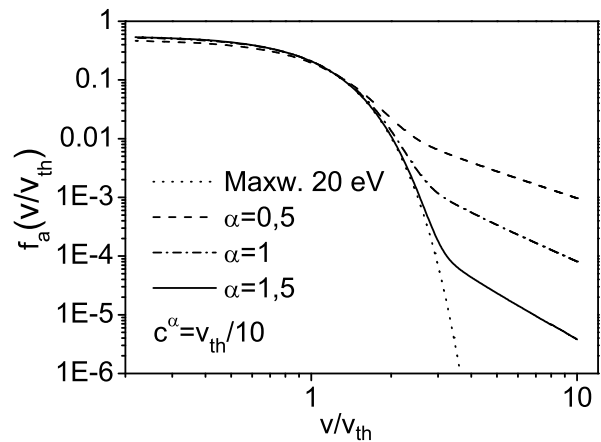} \caption{\label{fig:effVDFlevy}
Plot of the apparent velocity distribution corresponding to
L\'{e}vy velocity PDF in a logarithmic scale. The dotted line
corresponds to the Maxwellian which would be observed in the
absence of fluctuations. The velocity is plotted in units of the
thermal velocity $v_{th}$. The existence of an algebraic decay of
exponent $-\alpha+1$ in the tail of the apparent VDF is clearly
seen.}
\end{figure}

\noindent Such a behavior would for instance arise if the velocity
PDF were a L\'{e}vy distribution (see section \ref{sec:five}).
Examples of these distributions are plotted on Fig.
\ref{fig:levyvelocity}, for $\alpha=0.5, \ 1,\ 1.5$ and
$c^{\alpha}=v_{th}/10$. The resulting apparent velocity are
plotted on Fig. \ref{fig:effVDFlevy}, and exhibit a power-law
behavior in their tails.
However, it should be pointed out that for this effect to be
observable, large amplitude velocity fluctuations of the order of
a few thermal velocity $v_{th}$ should actually occur. Finally, it
should be noted that in a magnetized plasma, the physics
underlying parallel and perpendicular  velocity fluctuations are
different. The latter are related to the electric field
fluctuations through

\begin{equation}
  \textbf{u}_{\bot}\simeq\frac{\textbf{E}\times\textbf{B}}{B^{2}}
,
\end{equation}

\noindent whereas the former can arise from a Kelvin-Helmholtz
like instability associated to the existence of a perpendicular
gradient of parallel velocity. Changing the orientation of the
line of sight would allow to investigate each of these different
cases.

\subsection{Temperature fluctuations}

Finally, we consider the case where only the ion temperature
fluctuates, and for which the apparent VDF reads

\begin{equation}\label{eq:profilTemp}
  f_{a}(v_{z})=\int_{0}^{+\infty}W(T) \ f(v_{z},T)dT .
\end{equation}

\begin{figure}
\epsfig{scale=0.8 ,file=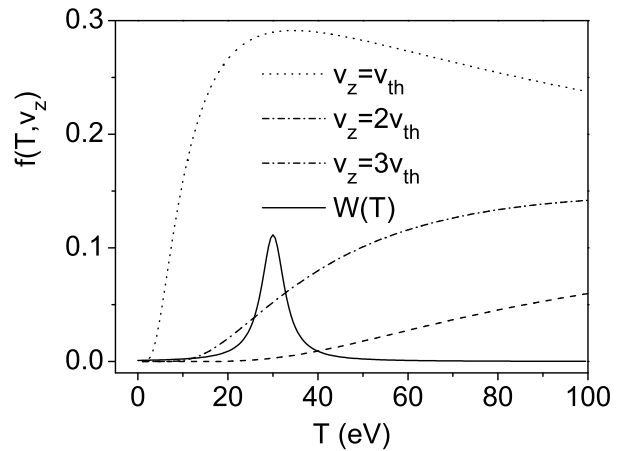} \caption{\label{fig:integrand}
Plot of the local VDF $f(T,v_{z})$ as a function of $T$ for three
different values of the component of the velocity along the LOS
$v_{z}=v_{th}$, $2v_{th}$, $3v_{th}$, where $v_{th}$ is the
thermal velocity for 30 eV. A model distribution $W(T)$, peaked
around $T_{0}=30$ eV is also plotted (solid line). As $v_{z}$ is
increased, the contribution of $T_{0}$ in the calculation of the
apparent VDF becomes all the more negligible than the tail of
$W(T)$ decreases slowly.}
\end{figure}

\noindent The latter is not a convolution product, in opposition
to the case of velocity fluctuations. To begin with, the apparent
temperature defined by Eq. (\ref{eq:Teff}), is given by

\begin{equation}
  T_{a}=\int_{0}^{+\infty}dT \ W(T) \ T ,
\end{equation}

\noindent and is thus equal to the mean temperature of the
distribution $W(T)$. Hence, $T_{a}$ does not depend on the
temperature fluctuations variance. The profile is obtained as a
weighted sum of Gaussians of different widths, and thus cannot
stay rigorously Gaussian itself. Nevertheless, for a sharp
temperature PDF peaked around $T_{0}$, the actual deviations from
Gaussianity should not be very important, as the dominant
contribution in the integral of Eq. (\ref{eq:profilTemp}) is
expected to come from the neighborhood of $T_{0}$. However, while
leading to accurate results for the central part of the profile,
this line of argument is not correct for the line wings. Indeed,
the value of $f(T_{0},v_{z})$ scales with $v_{z}$ as

\begin{equation}
  f(T_{0},v_{z})\propto \exp\left(-\frac{v_{z}^{2}}{\xi
T_{0}}\right),
\end{equation}

\noindent and therefore strongly decreases as $v_{z}$ increases.
Consequently, as shown on Fig. \ref{fig:integrand}, the
contribution of the maximum of the temperature PDF in the integral
becomes negligible for large enough $v_{z}$ (i.e. in the wings of
the apparent VDF), and this especially if $W(T)$ has a slowly
decreasing tail.

\begin{figure}
\epsfig{scale=0.8 ,file=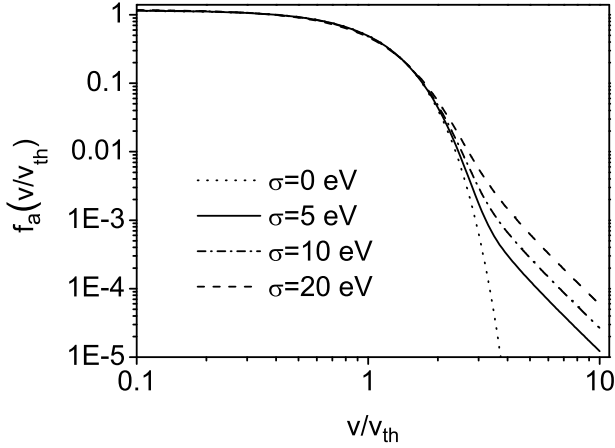}
\caption{\label{fig:sinaiprofile} Plot of the apparent VDF on a
logarithmic scale for $T_{0}=30$ eV, $k=10$ and for different
values of $\sigma=5,10,20$ eV. $v_{th}$ stands for the thermal
velocity for $T=30$ eV. These plot show the asymptotic power law
behavior. The value of the exponent is -3.2 here. The dotted line
corresponds to the Gaussian Doppler profile obtained for 30 eV.
The deviations from this gaussian profile becomes more and more
important as $\sigma$ grows.}
\end{figure}
For instance, an algebraic behavior for the temperature PDF
implies a similar one for the measured profile. The relation
between the exponents can be obtained in the following manner,
noting that for large velocities the apparent VDF can be
approximated by

\begin{equation}\label{eq:powerlawdemo}
  f_{a}(v_{z})\sim
  \int_{v_{z}^{2}/\xi}^{+\infty}W(T)\frac{1}{\sqrt{T}}dT .
\end{equation}


\noindent Using then a power-law ansatz for the temperature PDF,
the following result is readily obtained

\begin{equation}\label{eq:powerlawres}
  W(T)\propto\frac{1}{T^{\alpha+1}}  \Longleftrightarrow
f_{a}(v_{z})\propto\frac{1}{|v_{z}|^{2\alpha+1}}.
\end{equation}

\begin{figure}
\epsfig{scale=0.8 ,file=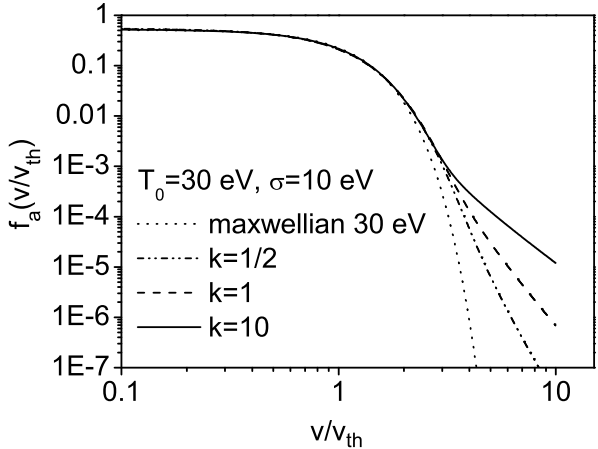}
\caption{\label{fig:apparentVDFk} Plot of the apparent VDF on a
logarithmic scale for $T_{0}=30$ eV, $\sigma=10$ eV and for
different values of $k$, $k=1,1/2,10$. $v_{th}$ again stands for
the thermal velocity for $T=30$ eV, and the dotted line
corresponds to the Gaussian Doppler profile obtained for 30 eV.}
\end{figure}

\noindent For example, let us consider the case in which the
temperature fluctuations PDF is the Sinai distribution given by
Eq. (\ref{eq:Tsallisinai}), and plotted in a logarithmic scale on
Fig. \ref{fig:sinaisigma} for $k=10$, $T_{0}=30$ eV, and for
different values of $\sigma$ ranging from 5 to 20 eV. The bulk of
the apparent VDF remains very close to that of the Maxwellian at
30 eV (dotted line) for every value of $\sigma$. However, the
discrepancies become important in the apparent VDF tails (i.e.
Doppler line wings), all the more so $\sigma$ is increased. In
addition, the tails are found to exhibit a linear behavior in
logarithmic scale, which signals a power-law dependence. The
exponent which characterizes this algebraic decay should take the
value $-3-2/k$ according to Eq. (\ref{eq:powerlawres}). The $k$
dependence can be checked on Fig. \ref{fig:apparentVDFk} where the
apparent VDF is plotted for $\sigma=10$ eV and for different
values of $k$ ($k=1,1/2,10$), i.e. different correlation
strengths. The stronger the correlations, the larger the
deviations from the Maxwellian.

In the frame of the Sinai model, the exponent $\alpha$
characterizing the apparent VDF power law decay is larger than 3.
Other turbulence models could lead to smaller exponents. Let us
indeed investigate the case in which the temperature PDF is a
L\'{e}vy distribution of indexes $0<\alpha<2$ and $-1<\beta<1$,
denoted by $\mathcal{L}_{\alpha,\beta}(T)$ \cite{Paul}. In the
Fourier space, one has

\begin{equation}
\ln
\tilde{L}_{\alpha,\beta}(k)=-c|k|^{\alpha}\left(1+i\beta\frac{k}{|k|}\omega(k,\alpha)\right)
,
\end{equation}

\noindent where $c$ controls the width of the distribution, and
the function $\omega(k,\alpha)$ is defined by

\begin{equation*}
\omega(k,\alpha)=
  \begin{cases}
    \tan(\pi\alpha/2) & \text{for } \alpha\neq 1, \\
    (2/\pi)\ln |k| & \text{for } \alpha=1.
  \end{cases}
\end{equation*}

\noindent For $0<\alpha<1$ and $\beta=-1$, $W(T)\equiv 0$ for
negative arguments, as should be the case for the temperature
field. The Fourier transform $\tilde{f}_{a}(k)$ of the apparent
VDF is given by

\begin{equation}
\tilde{f}_{a}(k)=\int_{0}^{+\infty}\mathcal{L}_{\alpha,-1}(T) \
\exp\left(-\frac{\xi T}{4}k^{2}\right)\ dT ,
\end{equation}

\noindent and can be calculated explicitly using the following
result \cite{Paul} which gives the Laplace transform of a L\'{e}vy
distribution

\begin{equation}
  \int_{0}^{+\infty}\mathcal{L}_{\alpha,-1}(T)\exp(-sT)dT=\exp{-c s^{\alpha}}
,
\end{equation}

\noindent with $s=k^{2}/2m+\imath 0$, m standing for the emitters
mass. The apparent VDF is thus found to be a symmetrical L\'{e}vy
distribution of indexes $\alpha'=2\alpha$ and $\beta'=0$

\begin{equation}\label{eq:VDFefflevy}
  f_{a}(v)=\frac{\sqrt{2m}}{c^{1/2\alpha}}\mathcal{L}_{2\alpha,0}\left(\frac{\sqrt{2m}}{c^{1/2\alpha}}v\right).
\end{equation}

\noindent Asymptotically,

\begin{equation}
  f_{a}(v)\sim \frac{1}{|v|^{2\alpha+1}} ,
\end{equation}

\noindent in accordance with Eq. (\ref{eq:powerlawres}). Here, the value of $\alpha$
is such that $1<2\alpha+1<3$ and therefore spans a different range than in the Sinai model.\\

In this idealized model where only temperature fluctuates, the
analysis of the apparent VDF tails, i.e. of the line wings, allows
to retrieve information on the statistical properties of
temperature fluctuations. Indeed, power law decaying tails would
for instance be a signature of a similar behavior for the
temperature PDF. In addition, in this case, an analysis of the
experimental value of the exponent would allow to distinguish
between different turbulence models, corresponding for example
either to a Sinai or a L\'{e}vy PDF.


%


\subsection{Discussion}

The study of the case where only one variable fluctuates leads to
several enlightening conclusions. First of all, the Doppler
profile is only affected by ion temperature and fluid velocity
fluctuations along the line of sight, in contrast to the line
brigthness which essentially reflects the variations of the
density. In addition, the bulk of the line appears to be weakly
sensitive to the presence of low frequency turbulence, unless the
velocity fluctuations variance becomes comparable to the thermal
velocity. Therefore, turbulence can indeed be neglected if we
restrict ourselves to the study of the core of the line, as is
usually done \cite{Kubo,Hey,Koubiti02}. Conversely, the Doppler
line wings behavior is significantly altered by turbulent
fluctuations having non-Gaussian PDF. More precisely, long tails
for the PDF translates into long tails for the apparent VDF, i.e.
slowly decreasing line wings. In this sense, modifications on line
wings are associated to intermittency. A comparison with
experimental spectra would require further work both from the
theoretical and experimental sides, and will not be attempted
here. In particular, a refined model should simultaneously retain
velocity and temperature fluctuations. Indeed, velocity and
temperature effects cannot be distinguished \textit{a priori}. In
fact, examples where fluctuations of both fields lead to a power
law behavior for line wings have been presented above. In
addition, the couplings between density, velocity and temperature
fluctuations, which are responsible for anomalous transport,
should also be taken into account. To include these effects in our
model, one could either rely on the determination of a joint PDF,
or resort to a numerical integration of the fluid equations, which
would allow a straightforward calculation of the apparent VDF from
Eq. (\ref{eq:fbarexpli}). From the experimental point of view,
line wings may seem difficult to measure, but it should be kept in
mind that the acquisition time can in principle be chosen as large
as needed. The only limitation here is the actual duration of the
discharge stationary phase during which the measurements are
performed.

\section{\label{sec:seven} Apparent non-Boltzmann statistics}





In the above section, we have shown that the apparent VDF may
significantly differ from the Maxwellian calculated using the
averaged fields. For the sake of simplicity, let us only consider
temperature fluctuations here. The fact that the apparent VDF can
be a L\'{e}vy distribution highlights a connection between
spectroscopy, turbulence and anomalous statistics involving
power-law tails, such as the L\'{e}vy statistics. Indeed, it
should be emphasized that in the case where no other observable
than the spectral line shape is available (e.g. in Astrophysics),
it is by no mean possible to determine whether the observed plasma
is actually turbulent or homogenous. Therefore, if the temperature
PDF is a L\'{e}vy distribution $\mathcal{L}_{\alpha,-1}(T)$, the
Doppler spectra might be interpreted as resulting from an
homogeneous and stationary plasma governed by Levy statistics. In
other words, everything happens as if the plasma under study were
in a non-equilibrium stationary state characterized by the
L\'{e}vy distribution of Eq. (\ref{eq:VDFefflevy}). This
stationary state can be seen as resulting from a relaxation
process governed by the following Fractional Fokker-Planck
Equation (FFPE) \cite{CNSNS,jespersen99,chechkin02}

\begin{equation}\label{eq:FFPE}
   \frac{\partial f_{a}(v,t)}{\partial t}=\bar{\nu}\frac{\partial}{\partial
  v}[v f_{a}]+\bar{D}\frac{\partial^{2\alpha}f_{a}}{\partial
  |v|^{2\alpha}} .
\end{equation}

\noindent where $\bar{\nu}$ and $\bar{D}$ are such that
$\bar{D}/\bar{\nu}=2\alpha c /(2m)^{\alpha}$. Here, the fractional
derivative is defined in the sense of Riesz \cite{Paul}

\begin{equation}
\frac{\partial^{2\alpha}f_{a}}{\partial
|v|^{2\alpha}}=TF^{-1}\left[-|k|^{2\alpha}\tilde{f}_{eff}\right] .
\end{equation}

\noindent The usual Fokker-Planck equation (FPE) is recovered for
$\alpha=1$. In our case $\alpha<1$, and the apparent VDF cannot be
Gaussian. The main physical difference between the FPE and the
FFPE given by Eq. (\ref{eq:FFPE}) is the spatial non locality of
the latter, obvious from the definition of the fractional
derivative. This non locality is a consequence of the existence of
flights connecting distant regions in the velocity space (the
so-called L\'{e}vy flights). This property can be traced back to
the underlying description of the turbulent plasma. Indeed, at the
microscopical scale the trajectory of the radiators can be
modelled by a Langevin equation with gaussian white noise
\cite{Beck01,Resibois}. This model describes the collisional
relaxation of the local velocity distribution toward the local
Maxwellian Eq. (\ref{eq:locmaxw}). Using the
fluctuation-dissipation theorem and the expression of the
diffusion coefficient stemming from a random walk model
\cite{Resibois} leads to

\begin{equation}
  \frac{\langle \Delta v^{2}\rangle}{\tau_{j}}\sim
\nu\frac{k_{B}T}{m},
\end{equation}

\noindent where $T$ is the local temperature and $\tau_{j}$ the
typical time between two jumps in the velocity space. Temperature
thus determines the characteristic size of jumps in the velocity
space. Therefore, high probabilities for large temperature
fluctuations in the actual turbulent plasma imply high
probabilities for flights in the apparent velocity space. This
provides a simple physical picture explaining why the temperature
PDF and the apparent VDF asymptotical behavior are linked, and
leads to a deeper understanding of Eq. (\ref{eq:powerlawres}).
 Our results are reminiscent of those presented in
references \cite{Wilk01,Beck01,Beck03}, where a similar
interpretation of Tsallis non extensive statistical mechanics
occurrence was proposed. The latter case arises if the temperature
PDF is such that $1/T$ is gamma distributed \cite{Wilk01}. Let us
emphasize that in our model, the temperature PDF shape is not
arbitrary. In fact, it has to be determined from the fluid
equation satisfied by the temperature field in the plasma under
consideration, in which relevant expressions for both the source
term and the flux have to be specified (see Eq. \ref{eq:fluidX}).
For each of these expressions, the non-linear character of the
latter equation should give rise to a different non-gaussian
statistical behavior, i.e. lead to a specific PDF, and therefore
to a particular apparent statistics. A natural extension of this
work would be to determine what properties fluid equations should
have so as to lead to a L\'{e}vy distribution for temperature.

\section{Conclusion and perspectives}

In this paper, we have presented a model retaining low frequency
turbulence in Doppler line shape calculations. This approach is in
particular relevant to the modelling of lines routinely measured
in edge plasmas of fusion devices. We have shown that in presence
of low frequency turbulence, a straightforward analysis of Doppler
profiles yields an apparent velocity distribution function. This
apparent VDF is a spatial and time average of the local VDF. To
investigate its shape, we have used a statistical description of
the plasma turbulent fluctuations, relevant whenever the
acquisition time of the spectrometer is large with respect to the
typical turbulent time scale. The resulting expression for the
apparent VDF involves the joint Probability Density Function of
the fluctuating fields. Next, considering the case where only one
variable fluctuates, we have obtained several new results. While
density fluctuations do not affect Doppler line shapes, velocity
or ion temperature fluctuations can strongly influence line wings.
This is especially the case when their PDF have long tails such as
power laws. It might therefore be possible to diagnose such a
behavior by the mean of line shapes, once Stark effect has been
carefully ruled out. A reliable comparison with experiments would
imply dedicated measurements which are not yet available, but also
further modelling. In particular, the use of a turbulence code
would be very helpful for diagnosis purposes, and this possibility
will be investigated in a future work. From a more fundamental
point of view, our work sheds light on some possible connections
between turbulence, spectroscopy and non Boltzmann statistics,
such as those involving L\'{e}vy or Tsallis distributions. Our
approach furthermore relates the occurrence of one of these
particular statistics to the properties of the fluid equations
describing turbulence. Our model thus provides a frame to
investigate both experimentally and theoretically some of the
fundamentals aspects of the statistical properties of the physical
observables in out of equilibrium plasmas.

\begin{acknowledgments}
The authors would like to thank F. B. Rosmej for helpful
discussions. This work is part of a collaboration (LRC DSM 99-14)
between the Laboratoire de Physique des Interactions Ioniques et
Mol\'{e}culaires and the D\'{e}partement de Recherches sur la
Fusion Contr\^{o}l\'{e}e, CEA Cadarache.
\end{acknowledgments}

%

\newpage 
\bibliography{marandet}

\end{document}